\documentclass[a4paper,12pt]{article}
\usepackage{epsfig,epsf,latexsym}
\begin{document}
%
%
%
%
%
%
%
%
%
\newcommand{\x}{\cdot}
\newcommand{\ra}{\rightarrow}
\newcommand{\pom}{\mbox{Pomeron}}
\newcommand{\flux}{\mbox{$F_{{I \!\! P}/p}(t, \xi)$}}
\newcommand{\fluxpap}{\mbox{$F_{{\cal P}/p}^{\bar{p}p}(t, \xi)$}}
\newcommand{\fluxgmp}{\mbox{$F_{{\cal P}/p}^{ep}(t, \xi)$}}
\newcommand{\ap}{\mbox{$\bar{p}$}}
\newcommand{\pap}{\mbox{$\bar{p} p$}}
\newcommand{\SPS}{\mbox{S\pap S}}
\newcommand{\xp}{\mbox{$x_p$}}
\newcommand{\xf}{\mbox{$x_F$}}
\newcommand{\sumet}{\mbox{$\Sigma E_t$}}
\newcommand{\mpr}{\mbox{${m_p}$}}
\newcommand{\mpi}{\mbox{${m_\pi}$}}
\newcommand{\rs}{\mbox{$\sqrt{s}$}}
\newcommand{\rsp}{\mbox{$\sqrt{s'}$}}
\newcommand{\rsps}{\mbox{$\sqrt{s} = 630 $ GeV}}\newcommand{\lum}{\mbox{$\int
{\cal L} {dt}$}}
\newcommand{\T}{\mbox{$t$}}
\newcommand{\abt}{\mbox{${|t|}$}}
\newcommand{\pp}{\mbox{$pp$}}
\newcommand{\hard}{\mbox{$\beta F_{g/I \!\! P}(\beta) \, = \, 6 \, \beta \,
(1-\beta)^1$}}
\newcommand{\soft}{\mbox{$\beta F_{g/I \!\! P}(\beta) \, = \, 6 \,
(1-\beta)^5$}}
\newcommand{\sigdifjets}{\mbox{$\sigma_{sd}^{jets}$}}
\newcommand{\sigpomjets}{\mbox{$\sigma_{{\cal P}p}^{jets}$}}
\newcommand{\sigdiftot}{\mbox{$\sigma_{\bar{p} p}^{\rm tot \, diff}$}}
\newcommand{\sigpomtot}{\mbox{$\sigma_{I \!\! P p}^{\rm tot}$}}
\newcommand{\sigpompom}{\mbox{$\sigma_{{\cal P} {\cal P}}^{\rm tot}$}}
\newcommand{\sigpptot}{\mbox{$\sigma_{pp}^{\rm tot}$}}
\newcommand{\sigpomzero}{\mbox{$\sigma_{{\cal P}p}^o$}}
\newcommand{\dsig}{\mbox{${d^2 \sigma }\over{d \xi dt}$}}
\newcommand{\alamb}{\mbox{$\overline{\Lambda^{\circ}}$}}
\newcommand{\lamb}{\mbox{$\Lambda^{\circ}$}}
\newcommand{\pt}{\mbox{$P_t$}}
\newcommand{\PRET}{\mbox{\Proton-\sumet}}
%
\newcommand{\xpom}  {\mbox{$x_{I \! \! P}$}}
\newcommand{\pmm}         {\mbox{${\cal P}$}}
\newcommand{\gm}         {\mbox{$\gamma^{*}$}}
\newcommand{\gmp}         {\mbox{$\gamma^{*} p$}}
\newcommand{\siggp}      {\mbox{$\sigma_{\gamma^{*} p}^{\rm tot}$}}
\newcommand{\siggpm}     {\mbox{$\sigma_{\gamma^{*} {\cal P}}^{\rm tot}$}}
\newcommand{\FtwoDtwo}       {\mbox{$F_2^{D(2)}$}}
\newcommand{\FtwoDthree}       {\mbox{$F_2^{D(3)}$}}
\newcommand{\xpomFtwo}     {\mbox{$\xi \FtwoDthree$}}
\newcommand{\qsq}        {\mbox{$Q^2$}}
\newcommand{\mx}         {\mbox{$M_X$}}
\newcommand{\mxsq}       {\mbox{$M_X^2$}}
\newcommand{\w}          {\mbox{$W$}}
\newcommand{\wsq}        {\mbox{$W^2$}}
\newcommand{\fluxint}    {\mbox{$f_{{\cal P}/p}(\xi)$}}
\newcommand{\eps}        {\mbox{$\epsilon$}}
\newcommand{\alf}        {\mbox{$\alpha '$}}
\newcommand{\xiP}         {\mbox{$\beta$}}
\newcommand{\xip}         {\mbox{$\xi_{p}$}}

\begin{titlepage}
\vspace{4cm}
\begin{flushright}

{23 December, 2003}\\
CERN TH/2003-232\\
UCLA EPP/2003-101\\
\end{flushright}
\vspace{4ex}
\begin{center}
\LARGE
{\bf \boldmath Central Higgs Production at LHC}\\
{\bf \boldmath from Single--${I \!\! P}$omeron--Exchange}\\

\normalsize
\vspace{9 ex}
Samim Erhan$^a$, Victor T. Kim$^{bc}$ and Peter E. Schlein$^a$ \\
\vspace{3.0mm}
$^a$University of California$^{*}$, Los Angeles, California 90095, U.S.A. \\
$^b$St. Petersburg Nuclear Physics Institute, Gatchina 188300, Russia \\
$^c$CERN, CH-1211, Geneva 23, Switzerland\\

\end{center}
\vspace{11 ex}
\begin{abstract}

Contrary to common perceptions about systems produced in
Single--\pom --Exchange (SPE) $pp$ interactions, the hard diffractive process
discovered at the CERN \SPS --Collider leads to
dominant central production of Higgs bosons at the LHC.
The rate for SPE production of Higgs bosons is
calculated to be 7-9\% of the total inclusive Higgs rate.
In addition, an SPE measurement program of dijet events is outlined for the
early days of LHC
running which should answer many fundamental questions about the \pom\
structure and its effective flux factor in the proton.
\end{abstract}
\vspace{3cm}
\begin{center}
Submitted to European Physical Journal C
\end{center}
\vspace{4cm}
\rule[.5ex]{16cm}{.02cm}
$^{*}$\ Supported by U.S. National Science Foundation Grant PHY-9986703\\
\end{titlepage}
\setlength{\oddsidemargin}{0 cm}
\setlength{\evensidemargin}{0 cm}
\setlength{\topmargin}{0.5 cm}
\setlength{\textheight}{22 cm}
\setlength{\textwidth}{16 cm}
\setcounter{totalnumber}{20}
\clearpage\mbox{}\clearpage
\pagestyle{plain}
\setcounter{page}{1}
\section{Introduction}
\label{sect:intro}
\indent

A primary physics goal of the Large Hadron Collider project (LHC) is the
discovery and
study of the Higgs boson, whose production is expected to be mainly via the
gluon--fusion process, $gg \ra  Higgs$.
The data analysis strategy for such interactions has long been an important
topic in planning for the major experiments, ATLAS and CMS.
Although certain types of background are irreducible, for example
$gg \ra b \bar{b}$, it nonetheless seems self evident that a general reduction
of hadronic activity in the final state must improve the capability for Higgs
detection and isolation.

In a typical interaction in which a 120 GeV Higgs boson is produced at the LHC,
a gluon participating in the fusion process has only about 1\% of the 7 TeV beam
momentum.
Thus, about 99\% of the beam energy resides in other partons which are
responsible for the overall hadronic activity.
An alternate approach to inclusive Higgs study is to focus on the
\pom --Exchange class of events (also known as ``diffraction''), where much of
the excess beam energy is carried away by a single proton on one or both sides
of the interaction.
Recently, much attention has been paid to Higgs production via
Double--\pom -Exchange \cite{nacht,khoze}
In this paper, because the cross section is much larger, we discuss the
Single--\pom --Exchange  process (SPE) where only a single proton
emerges.
\begin{equation}
p \, + \, p \, \, \ra \, \, p \, + \, X \, \, \ra \, \,  p \, + \, (H  Y),
\label{eq:difhiggs}
\end{equation}
The final state proton and the ($H \, Y$) system each recoil with
more than 90\% of the initial 7 TeV beam momentum.
Previous discussions of this reaction are in \cite{veneziano,stirling,ingelman}
with cross section predictions ranging from 1\% to 25\% of the total Higgs cross
section.

We now understand that diffractive processes occur because of the existence of
colorless gluon-dominant clusters (for example, di--gluons) in the partonic sea
of a proton which interact with the other beam proton.
The existence of objects in a proton's sea with most likely momentum fraction
of their host proton near zero is linked to the observations starting about 40
years ago \cite{gellert} of inclusively--measured final--state protons from
inelastic interactions which possessed nearly all the initial beam momentum.

There is solid empirical evidence for the existence of such clusters.
The UA8 Experiment \cite{ua8dijets} at the CERN \SPS --Collider observed jet
production in this class of events, demonstrating the hard
partonic structure of the clusters.
The H1 experiment \cite{h1gluons} at the DESY HERA $ep$-Collider studied
$\gamma^*$ interactions with rapidity gaps and demonstrated that the
clusters have dominant gluon structure.

In the present paper, we show that the hard \pom --proton collision which
creates the ($H \, Y$) system in React.~\ref{eq:difhiggs} results
in the Higgs emerging in the central region of the $pp$ center--of--mass.
To calculate the cross section for React.~\ref{eq:difhiggs}, we assume effective
factorization between the probabilities to find a \pom\ and
its interaction with the other proton \cite{es3}:
\begin{equation}
{{d^2\sigma}\over{dt d\xi}} \, \, = \, \, \flux \, \cdot \,
\sigma_{I \!\! P p \, \ra \, H \, Y} (M_X)
\label{eq:sighiggsx}
\end{equation}
where \flux\ is the \pom\ flux factor in the proton, which depends on the
squared 4--momentum transfer, $t$, to the proton and $\xi = 1 - \xp$, the
\pom 's momentum fraction of its host proton.
$\sigma_{I \!\! P p \, \ra \, H \, Y} (M_X)$ is the Higgs cross
section, where $M_X$ is the invariant--mass of the \pom -proton (or $H Y$)
system.
To good approximation, $M_X^2 = \xi s$, where $s$ is the total squared
c.m. energy.
We discuss the limitations of the cross section calculation and outline how a
series of measurements can be made at the LHC to test the fundamental aspects of
\pom --exchange reactions.

Section~\ref{sect:kine} contains a discussion of the kinematics of
React.\ \ref{eq:difhiggs} and shows what can be learned from the data without
prior knowledge of the \pom\ flux factor and its structure function.
Section~\ref{sect:sighiggs} contains the calculation of the \pom --proton
cross section in Eq.~\ref{eq:sighiggsx} and explains how the \pom\ structure
function can be determined in hadronic interactions.
Section~\ref{sect:flux} introduces the \pom\ flux factor and how it is
determined empirically.
Section~\ref{sect:total} then uses it to calculate the total cross section for
React.\ \ref{eq:difhiggs}.
Discussion and conclusions are in Sect.\ \ref{sect:conclude}.

\section{\boldmath Kinematics}
\label{sect:kine}
\indent

In the $pp$ center-of--mass (the laboratory system), the mass of the Higgs,
$M_H$, and its Feynman--$x$, $x_H^{pp}$, can be expressed in terms of the two
interacting gluons' momentum fractions of their host protons, $\xi_1$ and
$\xi_2$, respectively:
\begin{equation}
M_H^2 \, \, = \, \, \xi_1 \, \xi_2 \, s
\label{eq:mh}
\end{equation}
\begin{equation}
x_H^{pp} \, \, = \, \, \xi_1 \, - \, \xi_2,
\label{eq:xf}
\end{equation}
where $s = (14$~TeV$)^2$ at the LHC.
If $\xi_1$ is assumed to arise from an intermediate di--gluon state
(the \pom ), then:
\begin{equation}
\xi_1 \, \, = \, \, \beta \, \xi
\label{eq:xi1}
\end{equation}
where $\xi = 1 - \xp$ is the \pom 's momentum fraction of its host proton and
$\beta$ is the gluon's momentum fraction in the \pom .
Eqs.~\ref{eq:mh}, \ref{eq:xf} and \ref{eq:xi1} are combined to give:
\begin{equation}
x_H^{pp} \, \, = \, \, \beta \, \xi \, \, - \, \, {{M_H^2}\over{\beta \, \xi \,
s}}
\label{eq:higgsxf}
\end{equation}

\begin{figure}[t]
\begin{center}
\mbox{\epsfig{file= 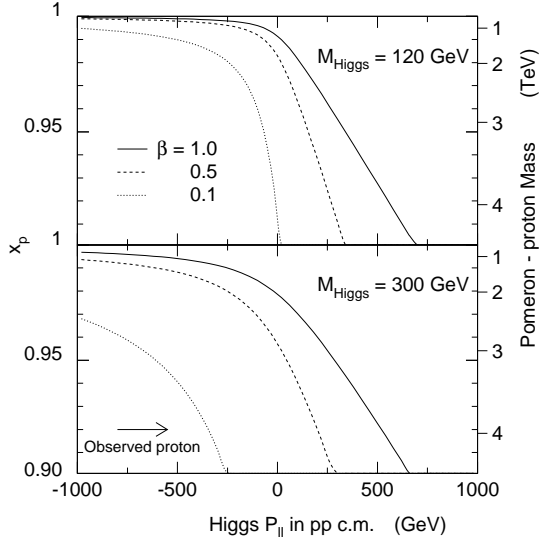,width=76mm}}
\end{center}
\caption[]{\em{
Higgs SPE kinematics at LHC for $M_H$ = 120 and 300 GeV,: Longitudinal Higgs
momentum in the laboratory vs.\ observed proton Feynman-\xp\ for three values of
a gluon's momentum fraction, $\beta$, of its host \pom .
Positive Higgs' momentum corresponds to the direction of the observed final
state proton.
}}
\label{fig:hkine}
\end{figure}

Figure~\ref{fig:hkine} displays Eq.~\ref{eq:higgsxf} for
React.~\ref{eq:difhiggs}.
We see that, for fixed $M_X$ (and therefore also $\xi$), the longitudinal Higgs
momentum in the laboratory is uniquely related to a value of $\beta$.
In other words, the observed distribution of $P_{||}$ reflects the \pom\
structure function.
The precise relationship also depends on the gluon structure function of the
proton, as discussed in the next section.
To sum over all possible $M_X$ values, knowledge of the \pom\ flux factor in the
proton is needed, as discussed in Sect.~\ref{sect:flux}.

We also note that, for fixed $M_X$ and for the maximum value, $\beta = 1$,
$P_{||}$ has a maximum value in the laboratory.
Fig.~\ref{fig:hkine} also shows that the centrality of Higgs production can be
influenced by an appropriate selection of \xp .
Furthermore, for $P_{||} = 0$, and if we require $\xi < 0.1$ (in order to have
relatively clean \pom\ exchange), there is a minimum value of $\beta$ that is
sampled; it is $\beta \sim 0.1$ for $M_H = 120$~GeV at the LHC and increases
either with larger mass or smaller $s$.


\section{\boldmath Higgs cross section in \pom --proton collisions}
\label{sect:sighiggs}
\indent

The basis of all cross sections given in this paper is
the cross section for gluon-gluon fusion into Higgs boson at a gluon-gluon
mass-squared, $\hat{s}$, which can be written as:
\begin{equation}
\sigma_{gg \ra H}(M_H, \hat{s}) \, \, = \, \,
K_{HO} \cdot \sigma_{LO} \cdot [\hat{s} \cdot \delta(\hat{s}-M_H^2)]
\label{eq:sighiggs}
\end{equation}
where the leading order Higgs cross section is \cite{Georgi:1977gs}:

\begin{equation}
\sigma_{LO} \, \, = \, \, \frac{G_F \, \, \alpha_S^2}{288 \, \sqrt{2} \, \pi} \,
\,
\Biggl| \frac{3}{4} \sum_q A_q(M_H/M_q) \Biggr|^2
\end{equation}
$A_q$ is the matrix element for the $gg\rightarrow H$ process,
dominantly via a $q$-quark loop in leading order. The last factor
in Eq.~\ref{eq:sighiggs} is the zero-width approximation for the
Higgs boson which, for $M_H \leq 400$ GeV, is known to be a good
approximation. The factor, $K_{HO}$, accounts for higher-order
perturbative QCD corrections and is taken in the
next-to-next-to-leading order \cite{KHO1}: $ K_{HO} = K_{NNLO}
\simeq 2 \, (3) $ for LHC (Tevatron) energy.

In the \pom --proton center-of-mass system with fixed invariant mass, $M_X$, we
write the cross section for inclusive Higgs boson production in terms of the
standard gluon distribution function of the proton \cite{pstruct} taken at the
scale $Q^2 = M_H^2$, the gluon distribution function of the \pom\ (see below)
and the cross section for $g g \ra H$ using Eq.~\ref{eq:sighiggs}:
\begin{equation}
\sigma_{\it I \!\! P p \ra H}(M_H, M_X) \, \, = \, \, C_g \cdot \int_{0}^{1}
\int_{0}^{1}
F_{g / p}(\xip) \cdot F_{g /\it I \!\! P}(\xiP)
\cdot \sigma_{gg \ra H}(M_H, \hat{s})
\, \, \, d\xiP \, d\xip
\label{eq:pompsig1}
\end{equation}
where:
\begin{equation}
\hat{s} \, = \, \xiP \, \xip \, M_X^2
\label{eq:shat}
\end{equation}
is the squared invariant mass of the 2-gluon system and \xiP\ and \xip\ are the
gluon momentum fractions of their host \pom\ and proton, respectively.
$C_g = 0.8$ is the estimated fraction of the \pom\ momentum that resides in the
gluons \cite{h1gluons}.

Equation~\ref{eq:shat} and the $\delta$-function in Eq.~\ref{eq:sighiggs} allow
Eq.~\ref{eq:pompsig1} to be simplified to:
\begin{equation}
 \sigma_{\it I \!\! P p \ra H}(M_H, M_X) \, \, = \, \, C_g \cdot K_{HO} \cdot
\sigma_{LO} \cdot \int_{\tau}^1
F_{g/p}(\frac{\tau}{\xiP}) \cdot F_{g/\it I \!\! P}({\xiP}) \cdot
\frac{\tau}{\xiP} \, \, d{\xiP}
\label{eq:pompsig2}
\end{equation}
where:  $\tau = (M_H/M_X)^2$.

The Feynman--x of a produced Higgs boson in the \pom --proton
center--of--mass is given in terms of the gluon momentum fractions as:
\begin{equation}
x_H \, \, = \, \, \xiP \, - \, \xip \, \, = \, \, \xiP \, - \, \tau/\xiP
\label{eq:xH}
\end{equation}
Equation~\ref{eq:xH} allows us to write the differential cross section in $x_H$
as:
\begin{equation}
\frac{d\sigma_{\it I \!\! P p \ra H}(M_H, M_X)}{dx_H} \, \, = \, \, C_g \cdot
K_{HO} \cdot \sigma_{LO} \cdot
F_{g/p}(\frac{\tau}{\xiP}) \cdot F_{g/\it I \!\! P}({\xiP}) \cdot
\frac{\tau}{\xiP} \cdot
\biggl[ 1 + \frac{x_H}{ (x_H^2 + 4\tau)^{1/2} } \biggr]
\label{eq:dsigdx}
\end{equation}
and where $\xiP = 0.5 \, \biggl[x_H + {(x_H^2 + 4 \tau)}^{1/2}\biggr]$.

\vspace{2mm}

\begin{figure}[htb]
\begin{center}
\mbox{\epsfig{file= 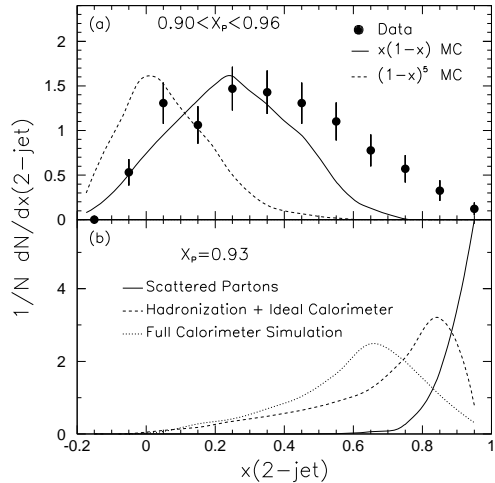,width=76mm}}
\end{center}
\caption[]{\em{
UA8 results \cite{ua8dijets} in the \pom --proton center--of--mass:
(a) Observed Feynman-\xf\ distribution of dijet systems.
The positive x-axis is in the \pom\ hemisphere;
(b) Results of $x$(2-jet) calculation in PYTHIA, assuming the
entire momentum of the \pom\ participates in the hard scattering as a gluon.
See discussion in text.
}}
\label{fig:hua8}
\end{figure}

We evaluate these equations for several possible \pom\ structure
functions \cite{is}.
Figure~\ref{fig:hua8}, reproduced from the original UA8 dijet paper
\cite{ua8dijets}, shows their experimental distribution of Feynman--$x$ for the
dijet system in the \pom --proton center--of--mass, ``$x$(2--jet)".
Almost all dijet systems travel in the \pom\
hemisphere\footnote{There is no major acceptance loss in the first part of the
proton hemiphere.}.
As pointed out in \cite{is}, such an observation means
that the \pom\ structure is much harder than the proton structure.
The two curves in Fig.~\ref{fig:hua8}(a) show the shapes of the expected
$x$(2--jet) distributions (with arbitrary normalization) for two examples of
possible \pom\ structure functions:
\begin{equation}
``Soft": \, \, \, \, \, \, \, \, \, \, \soft
\label{eq:soft}
\end{equation}
\begin{equation}
``Hard": \, \, \, \, \, \, \, \, \, \, \hard
\label{eq:hard}
\end{equation}
For both curves is Fig.~\ref{fig:hua8}(a), PYTHIA \cite{pythia1} was used to
calculate hadronization effects and initial--state and final--state radiation;
this was followed by a full simulation of the UA2 calorimeter \cite{ua2} which
detected the jets with $E_T > 8$~GeV.
This method of comparing ``smeared theoretical predictions" with observations is
a reliable way to compare theory with experiment  (see also comments in
\cite{chpww}).
About 30\% of the observed $x$(2--jet) distribution is harder than
the hard function of Eq.~\ref{eq:hard}.
UA8 therefore called the \pom\ structure ``Super-Hard'' \cite{ua8dijets}.
To parametrize this component, they assumed that the entire momentum
of the \pom\ participated in the hard scattering as a gluon.
In that case, $\xiP = 1$ and $x$(2--jet) = 1 -- \xip\ is the scattered
parton distribution before hadronization and detector effects; shown as
the solid curve in Fig.~\ref{fig:hua8}(b)\footnote{We note that because the
di--gluon mass is not unique for jet formation, the solid curve in
Fig.~\ref{fig:hua8}(b) displays the features of the gluon component of the
proton structure function}.
The dotted curve shows the effects of hadronization and full detector
simulation.
When added to the solid (Hard) curve in Fig.~\ref{fig:hua8}(a), the dotted curve
accounts nicely for the excess of events at large $x$(2--jet) in the
experimental distribution.

We therefore assume a \pom\ structure function to be the following
weighted sum of a gluonic Hard structure, and a smeared
$\delta$--function\footnote{For discussion of a
$\delta$--function component in the \pom , see \cite{cfs}.}. For
the latter, we use the function suggested by Alvero et al.
\cite{alvero}:
\begin{equation}
``Hard + Super Hard": \, \, \, \,  \, \, \, \, \, \, \, \, \,
\beta F_{g/\it I \!\! P}(\beta) \, = \,
0.7 \, \biggl[6 \, \beta \, (1 - \beta)\biggr] \, + \, 0.3 \, \biggl[19.8 \,
\beta^8  \,
(1 - \beta)^{0.3}\biggr]
\label{eq:hardSH}
\end{equation}
The essential property of this type of function required by the UA8, H1 and ZEUS
data \cite{ua8dijets,h1gluons,zeus1} is that it extends to $\beta = 1$.
The details of its shape do not greatly influence the Higgs predictions
presented here.
For all three types of \pom\ structure, their integrals are unity; in other
words, a momentum sum rule is assumed to be satisfied and the constituents of
the \pom\ carry all its momentum.
The momentum sum rule is discussed further in Sect.~\ref{sect:conclude}.
\begin{figure}[htb]
\begin{center}
\mbox{\epsfig{file= 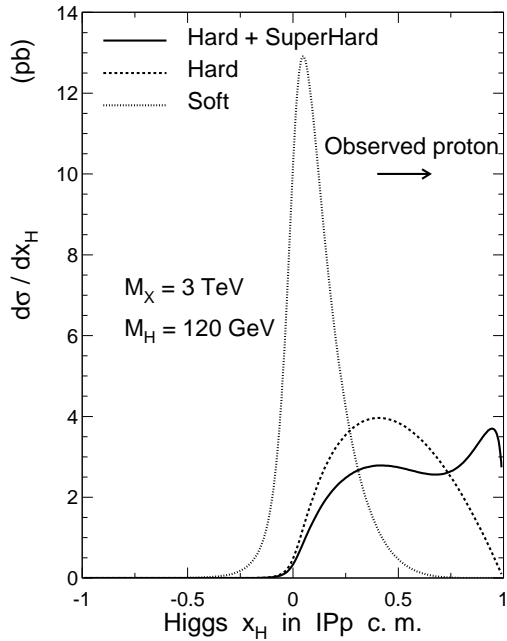,width=70mm}}
\end{center}
\caption[]{\em{ Differential cross section vs.\ Feynman-\xf\ of a
120 GeV mass Higgs boson in the \pom --proton center--of--mass in
React.~\ref{eq:difhiggs}, calculated assuming $M_X = 3$~TeV and
the three \pom\ structure functions given in Eqs.~\ref{eq:soft},
\ref{eq:hard} and \ref{eq:hardSH}. Positive $x_{H}$ is in the
observed proton direction. }} \label{fig:hpomp}
\end{figure}
The evaluation of Eq.~\ref{eq:dsigdx} using the \pom\ structure functions in
Eqs.~\ref{eq:soft}, \ref{eq:hard} and \ref{eq:hardSH} leads to the curves shown
in Fig.~\ref{fig:hpomp} for fixed $M_X = 3$~TeV and $M_H = 120$~GeV.

Following the arguments of \cite{is}, plots of experimental distributions
as in Figs.~\ref{fig:hua8} and \ref{fig:hpomp} directly compare the relative
hardness of the \pom\ structure with the proton structure.
With more common final states at the LHC, such as $t \bar{t}$, $b \bar{b}$ and
dijets it will in practice be possible to perform high  statistics
determination of the \pom\ structure function as a function of $Q^2$.

\begin{figure}[h]
\begin{center}
\mbox{\epsfig{file= 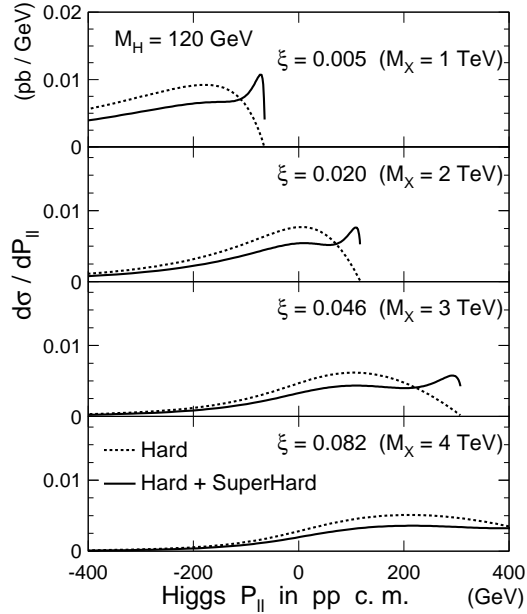,width=70mm}}
\end{center}
\caption[]{\em{
Higgs momentum along the beam axis in the laboratory frame (positive values are
in the observed proton direction) for four values of $M_X$ and $\xi$.
$M_H$ = 120 GeV.
The upper limit in Higgs momentum for each set of curves corresponds to
$\beta =1$, as seen in Fig.~\ref{fig:hkine}.
}}
\label{fig:hell4}
\end{figure}

Of course, following the arguments of Sect.~\ref{sect:kine}, it is possible to
also obtain the \pom\ structure function from the longitudinal momentum
distributions in the laboratory using data at fixed $\xp = 1 - \xi$ values.
The results from Eq.~\ref{eq:dsigdx} and Fig.~\ref{fig:hpomp} can be transformed to
the laboratory frame using the Lorentz transformation:
\begin{equation}
\gamma = \frac{1 + \xi}{2\sqrt{\xi}}.
\label{eq:gamma}
\end{equation}
This leads to the distributions of longitudinal Higgs momentum in the laboratory
shown in Fig.~\ref{fig:hell4}.
Figure~\ref{fig:hell4} also explicitly shows that, by selecting recoil proton
momentum ranges using the Roman pots, the mean Higgs longitudinal momentum can
be ``tuned".

\section{\boldmath The \pom\ flux-factor}
\label{sect:flux}
\indent

In order to relate our calculated $\sigma_{I \!\! P p \ra H Y}$  to
observed cross sections, the probability to find a \pom\ in the proton at fixed
$\xi$ and $t$ must be included.
In this section we summarize our present empirical knowledge of the \pom\ flux
factor and, in the process, also point out its limitations and where new
measurements will help.

The \pom\ flux factor has been determined for hadronic interactions by using the
inclusive inelastic diffractive process:
\begin{equation}
p \, \,  \, \, + \, \, p \, \,  \, \, \ra \, \, \, \, p \, \,
+ \, \, X
\label{eq:dif}
\end{equation}
and its corresponding $\pap$ reaction.
Such analyses are based on the assumption that the observed cross sections
are the product of two factors, the \pom\ flux factor and the relevant
\pom --proton total cross section, $\sigma_{I \!\! Pp}(M_X)$, at invariant mass,
$M_X$.
For React.~\ref{eq:dif}, the measured cross section is written as:
\begin{equation}
{{d^2 \sigma_{p p}^{\rm dif}}\over{dt d\xi}} \, \,
= \, \,  \flux \, \, \x \, \, \sigma_{I \!\! Pp \ra X}(M_X).
\label{eq:factorhad}
\end{equation}

Refs.~\cite{ua8dif,es2} contain the results of fitting Eq.~\ref{eq:factorhad} to
all available data on React.~\ref{eq:dif} from the CERN Intersecting Storage
Rings \cite{albrow} to the CERN \SPS --Collider \cite{ua8dif,ua4}, covering the
energy range, $\sqrt{s}$ = 23~GeV to 630~GeV and over the extended ranges, $|t|
< 2$~GeV$^2$ and $\xi < 0.09$.

The self-consistent set of fits reported in \cite{ua8dif,es2} are
equivalently important as those made to the diffractive DIS events by the H1
\cite{h1gluons} and ZEUS \cite{zeus2} collaborations.
The fits show that factorization  between \pom\ flux and \pom\ cross section, as
embodied in Eq.~\ref{eq:factorhad}, describes all available data at low $\xi$.
This is all the more remarkable, since the empirical flux factors differ in $ep$
and $pp$ interactions because of their different effective \pom\ trajectory
intercepts \cite{es3};

All fits in \cite{ua8dif,es2} use the following standard Regge form for
the flux factor \cite{dlflux} and the \pom --proton total cross
section\footnote{Eq.~\ref{eq:fitsig} is the same function typically used to
fit the $s$--dependence of other total cross sections \cite{tot1,tot2}.}.
Except for the \pom\ Regge trajectory which we discuss below,
the fitted parameters in \cite{ua8dif} are shown:

\begin{equation}
\flux = K \cdot |F_1(t)|^2 \cdot e^{(1.1 \pm 0.2)t} \cdot \xi^{1-2\alpha (t)}
\label{eq:fitflux}
\end{equation}

\begin{equation}
K \, \sigpomtot (s') \, \, =
\, \, (0.72 \pm 0.10) \cdot [(s')^{0.10} +
(4.0 \pm 0.6) (s')^{-0.32}]\,\,\,\,\, {\rm mb \, GeV^{-2}}.
\label{eq:fitsig}
\end{equation}
where $s' = M_X^2$.
Equation~\ref{eq:factorhad} shows that, in performing fits to data,
the scale factor, $K$, of the flux factor can not be separated from a scale
factor, $\sigma_0$, of the cross section.
Therefore, the value of the fitted product, $K\sigma_0$, is given in
Eq.~\ref{eq:fitsig}.
With the exponents\footnote{In this formula, ``$s'$" stands for ``$s'/s_0"$,
where $s_0$ = 1~GeV$^2$.} of $s' = \xi s$ fixed at 0.10 and -0.32, respectively,
$K \sigpomtot (s')$ requires the presence of  both \pom\ exchange and Reggeon
exchange terms.
With $|F_1(t)|^2$ in Eq.~\ref{eq:fitflux} set equal to the Donnachie--Landshoff
\cite{dlflux} form factor\footnote{$F_1(t)={{4 m_p^2 - 2.8t} \over{4 m_p ^2 -
t}}\, \x \, {1\over{(1-t/0.71)^2}}$}, the additional exponential factor is
required.

Finally, concerning $\alpha (t)$, Ref.~\cite{es2} demonstrated that, contrary to
Regge theory, in which Regge poles are fixed, the data on React.~\ref{eq:dif}
demonstrate that $\alpha (t)$  is an ``effective trajectory''.
That is, it changes with interaction energy. Its intercept at $t=0$ decreases
and its slope flattens as energy increases.
As predicted by Kaidalov et al. \cite{kaidalov}, the reason for this is likely
the result of complications due to multi--\pom --exchange effects.
In this connection, we note that the highest energy at which a fitted effective
trajectory has been obtained \cite{es2} is at $\sqrt{s} = 630$~GeV:
\begin{equation}
\alpha(t) \, = \, 1 \, + \, \epsilon \, + \, \alpha ' \, t \, + \, \alpha '' \,
t^2 \, = \, 1.035 \, + \, 0.165 \, t \, + \, 0.059 \, t^2
\label{eq:traj}
\end{equation}
It is shown in \cite{es2} that this effective trajectory is also compatible with
the published function which was fit to the CDF data
\cite{cdf} at $\sqrt{s} =  2$~TeV.

\begin{figure}[htb]
\begin{center}
\mbox{\epsfig{file= 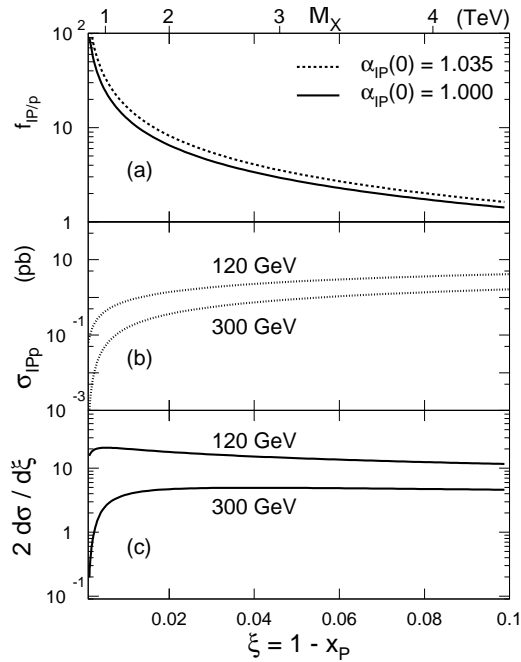,width=70mm}}
\end{center}
\caption[]{\em{
Dependence on $\xi$ and $M_X$ of:
(a)
Flux factor in Eq.~\ref{eq:fitflux} integrated over all $t$, using the \pom\
effective trajectory in Eq.~\ref{eq:traj} with $\epsilon$ = 0.035, and then 0.0.
$K$ = 0.78 GeV$^{-2}$ is assumed.
(b)
The \pom --proton to Higgs cross section, calculated for the two indicated Higgs
masses using Eq.~\ref{eq:pompsig2} and the Hard + SuperHard \pom\ structure
function.
(c)
Cross section for React.~\ref{eq:difhiggs} (the product of the solid curve in
(a) and each curve in (b)), multiplied by a factor of 2 to account for final
state proton detection in both arms.
}}
\label{fig:h3}
\end{figure}

Since we have no information as to the effective values of $\alpha (t)$ at
the LHC energy, all cross section calculations reported in this paper are given
for two values of the intercept, $\alpha(0)$ = 1.035 and 1.000.
The latter value was suggested by Schuler \& Sj{\"o}strand \cite{schsj} for use
at ultra--high energies.
Since all evidence shows that the effective \pom\ trajectory intercept decreases
with increasing energy, it seems like a reasonable choice to make.
In any case, our results do not depend greatly on which of these two values is
used.

Figure~\ref{fig:h3}(a) shows the $\xi$--dependence of the flux factor of
Eq.~\ref{eq:fitflux} integrated over all $t$.
The choice of the $K$ value assumed is perhaps the weakest aspect of the
predictions, for there is neither a measurement nor a reliable prediction for
$K$.
The best we can do is use the value of $K$ which Donnachie \& Landshoff
\cite{dlel} extracted from elastic-scattering data, namely
$K = 0.78$~GeV$^{-2}$, although given the observed $s$--dependence of the
effective \pom\ trajectory discussed above, we see no reason why this value of
$K$ should also apply precisely to inelastic diffraction where
multi--\pom --exchange effects may be different.
We therefore regard it as an approximation.
However, we discuss tests of this issue in Sect.~\ref{sect:conclude}.
%

\section{\boldmath Total Higgs cross section via SPE}
\label{sect:total}
\indent

\begin{figure}[htb]
\begin{center}
\mbox{\epsfig{file= 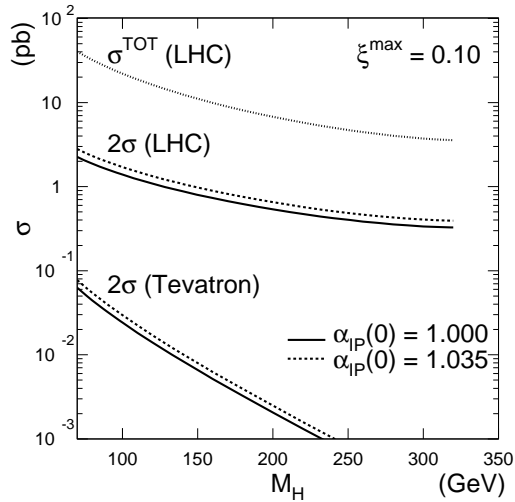,width=70mm}}
\end{center}
\caption[]{\em{
Upper curve is the total inclusive Higgs boson cross section vs.\ Higgs mass,
calculated for the gluon--gluon fusion process.
Central curves are the LHC cross section of Fig.~\ref{fig:h3}(c) integrated over
$\xi < 0.1$, for the two indicated values of $\alpha_{I \!\! P}(0)$.
The lower curves are the same, but for the Tevatron.
}}
\label{fig:htot}
\end{figure}

The total Higgs cross section via the SPE mechanism can now be calculated.
Fig.~\ref{fig:h3}(b) shows $\sigma_{I \!\! P p \ra H}(M_H, M_X) $ plotted vs.\
$M_X$ and $\xi$ for two values of $M_H$.
The product of the flux factor and $\sigma_{I \!\! P p \ra H}(M_H, M_X) $
is the $t$--integrated version of Eq.~\ref{eq:sighiggsx} and is shown in
Fig.~\ref{fig:h3}(c), multiplied by a factor of 2 to account for the use of
Roman pot spectrometers in both arms.
$\alpha_{I \!\! P}(0) = 1.0$ is assumed for both parts (b) and (c) of
Fig.~\ref{fig:h3}.
Finally, Fig.~\ref{fig:htot} shows the $\xi$--integral of the differential cross
section in Fig.~\ref{fig:h3} vs.\ Higgs mass.
At LHC, SPE Higgs production (with recoil protons measured in {\it either}
arm) is between 7\% and 9\% of the total Higgs cross section, depending on
$M_H$.

\begin{figure}[htb]
\begin{center}
\mbox{\epsfig{file= 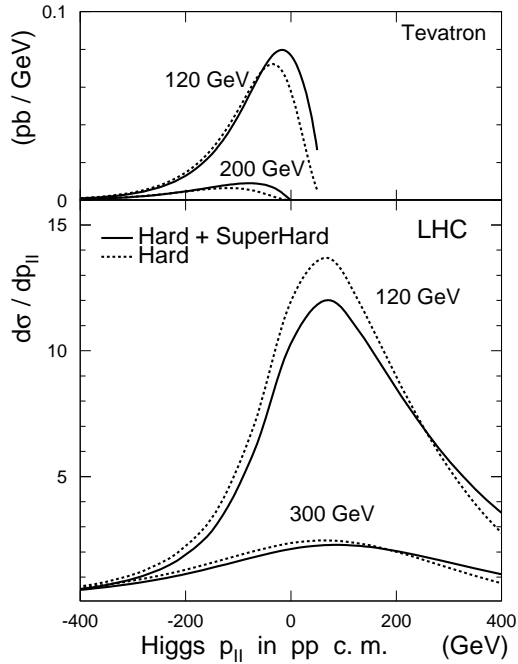,width=70mm}}
\end{center}
\caption[]{\em{
Differential cross sections at LHC and Tevatron for React.~\ref{eq:difhiggs}
vs.\ $P_{||}$ of the Higgs in the laboratory frame for the two indicated Higgs
masses and two assumed \pom\ structure functions.
}}
\label{fig:hell}
\end{figure}

The Higgs longitudinal momentum in the laboratory integrated over all $M_X$ is
shown in Fig.~\ref{fig:hell}.
This figure gives us global views of the centrality of Higgs production
integrated over all $\xi$ in the range: $0 < \xi < 0.1$.
The distributions do not depend greatly on the detailed shape of
the \pom\ structure function at large $\beta$.

\section{Discussion \& Conclusions}
\label{sect:conclude}
\indent

We have drawn attention to the method used by the UA8 Collaboration
\cite{ua8dijets,is} to determine the \pom\ structure function.
The longitudinal momentum distribution of the system produced via hard
scattering reflects the relative hardnes of the \pom\ and proton structure
functions. The analysis can be caried out either in the \pom\ proton c.m. or in
the laboratory although, in the later case, it must be done at fixed value of
$\xp = 1 - \xi$.
Progress in further understanding the \pom\ will be possible at the LHC by
making  high--statistics measurements of more common final states than those
with Higgs bosons, for example dijets, $t \bar{t}$ or $b \bar{b}$.

One of the parameters most poorly understood is the normalization, $K$, of the
\pom\ flux factor in the proton.
To further understand this issue, the following ratio can be measured over a
large range of $t$ and $\xi$:
\begin{equation}
{\cal R} \, \, = \, \, \frac{\Delta \sigma_{p p \ra p (dijet + Y)}}
{\Delta \sigma_{p p \ra p (X)}}
\label{ratio1}
\end{equation}
If factorization between flux factor and $I \!\! P p$ cross sections remains
valid, the ratio, ${\cal R}$, will be found to be
independent of $t$ at fixed $\xi$.
If that is the case, then the flux factor cancels out of the ratio and we have:
\begin{equation}
{\cal R} \, \, = \, \, \frac{K \Delta \sigma_{I \!\! P p \ra (dijet + Y)}}
{K \Delta \sigma_{I \!\! P p \ra (X)}}
\label{eq:ratio2}
\end{equation}
Numerator and denominator of Eq.~\ref{eq:ratio2} have been multiplied by $K$,
because the denominator is then a measurable quantity
(by fitting Eqs.~\ref{eq:fitflux}, \ref{eq:fitsig} and \ref{eq:traj} to the
data).
The numerator of the right--hand side of Eq.~\ref{eq:ratio2} can then be
extracted and compared with the calculated dijet cross section.

This comparison allows $K$ to be determined although, strictly speaking, a
possible violation of the momentum sum rule in the calculated dijet cross
section means that the quantity measured is actually ``$fK$", where $f < 1.0$ if
the sum rule is violated.
UA8 carried out this procedure \cite{ua8jetsig} with their limited statistics
dijet data.
For a pure gluonic \pom , they found:
\begin{equation}
fK \, \, = \, \, 0.38 \pm 0.13
\label{eq:fk}
\end{equation}
We have already pointed out that the Donnachie-Landshoff value,
$K = 0.78$~GeV$^{-2}$  which was extracted from elastic scattering data may not
be precisely applicable to inelastic diffractive data, because of different
multi--\pom --exchange effects.
Thus, if the momentum sum rule is valid, Eq.~\ref{eq:fk} is a measure of $K$
for inelastic diffraction.
On the other hand, it could reflect a product of the two effects.
In either case, taking Eq.~\ref{eq:fk} at face value would mean that all cross
sections presented in this paper are overestimated by a factor of $2.0 \pm 0.7$.
Studying $fK$ obtained in this way for different final states and over a wide
range of kinematics should shed additional light on its correct interpretation.

In conclusion, we have shown that Hard Diffraction Scattering with
Single--\pom --Exchange leads to Higgs production dominantly in the central
region.
The extent to which the decrease in hadronic activity characteristic of such
reactions will actually improve Higgs identification and isolation is best shown
with the aid of full event simulation using the PYTHIA software package
\cite{pythia2}.
Such studies are underway.
We have also briefly outlined an initial program for SPE production of dijet
events which should lead to major improvements in our knowledge of the \pom\
structure and its flux factor in the proton.

\section*{Acknowledgements}
\indent

We are greatly indebted to the CERN laboratory for their continued hospitality.
This analysis work was carried out at the CERN site, as was the UA8
experiment whose results provided much of the underlying inspiration.
One of us (V.T.K.) is supported in part by the Russian Foundation for Basic
Research (RFBR), the INTAS and the U.S. N.S.F.


\end{document}